\documentclass[prb,aps,amsmath,amssymb,reprint]{revtex4-1}
\usepackage{graphicx}
\usepackage{dcolumn}
\usepackage{bm}
\newcommand{\p}{$\%$}

\newcommand{\tfn}{$\mathrm{Fe_{4}N}$}

\newcommand{\muB}{$\mathrm{\mu_{B}}$}

\newcommand{\pn}{$\mathrm{R{_{N_2}}}$}

\begin{document}

\title{Influence of interface and microstructure on magnetization of epitaxial \tfn~thin film}
\author {Nidhi Pandey$^1$, S. P\"{u}tter$^2$, S. M. Amir$^2$, V. R. Reddy$^1$, D. M. Phase$^1$, J. Stahn$^3$, Ajay Gupta$^4$ and Mukul Gupta$^{1*}$}

\address{$^1$UGC-DAE Consortium for Scientific Research, University Campus, Khandwa Road, Indore 452 001, India}

\address{$^2$J\"{u}lich Centre for Neutron Science (JCNS) at Heinz Mair-Leibnitz Zentrum (MLZ), Forschungszentrum J\"{u}lich GmbH, Lichtenbergstr. 1, 85748
Garching, Germany}

\address{$^3$Laboratory for Neutron Scattering and Imaging, Paul Scherrer Institut, CH-5232 Villigen PSI, Switzerland}

\address{$^4$Amity Center for Spintronic Materials, Amity University, Sector 125, Noida 201 303,
India}
\address{$^*$Corresponding author email: mgupta@csr.res.in}
\date{\today}

\begin{abstract}
Epitaxial \tfn~thin films grown on lattice-matched LaAlO$_3$ (LAO)
substrate using sputtering and molecular beam epitaxy techniques
have been studied in this work. Within the sputtering process,
films were grown with conventional direct current magnetron
sputtering (dcMS) and for the first time, using a high power
impulse magnetron sputtering (HiPIMS) process. Surface morphology
and depth profile studies on these samples reveal that HiPIMS
deposited film has the lowest roughness, the highest packing
density and the sharpest interface. We found that the
substrate-film interface and the microstructure play a vital role
in affecting the magnetic properties of \tfn~films. La from the
LAO substrate and Fe from the film interdiffuse and forms an
undesired interface spreading to an extent of about 10-20\,nm. In
the HiPIMS process, layer by layer type growth leads to a globular
microstructure which restricts the extent of the interdiffused
interface. Such substrate-film interactions also affect the
electronic hybridization and magnetic properties of \tfn~films.
The magnetic moment ($M_s$) was compared using bulk,
element-specific and magnetic depth profiling techniques. We found
that $M_s$ was the highest when the thickness of the interdiffused
layer was lowest. And such conditions can only be achieved in the
HiPIMS grown samples. Presence of small moment at the N site was
also evidenced by element-specific x-ray circular dichroism
measurement in HiPIMS grown sample. A large variation in the $M_s$
values of \tfn~films found in the experimental works carried out
so far could be due to such interdiffused layer which is generally
not expected to form in otherwise stable oxide substrate at low
substrate temperature $\approx$ 675\,K. In addition, a consequence
of substrate-film interdiffusion and microstructure results in
different kinds of different kind of magnetic anisotropies in
films grown using different techniques. A detailed investigation
of substrate-film interface and microstructure on the
magnetization of \tfn~film is presented and discussed in this
work.

\end{abstract}

\maketitle

\section{Introduction}
\label{1}

Ferromagnetic electrodes materials with large spin polarization
ratio (SPR) of transport electrons are of great interest for the
application development of spintronic devices. Numerous materials
have been actively investigated in this
context.~\cite{CrO2_SPR_PRB_2001, Heusler_SPR_APL_2014} Among
them, \tfn~recently elicited renewed interest as it is predicted
to show half-metallicity (spin polarization ratio $\approx$
100\p),~\cite{PRB06_Fe4N_SPR_100p_Kokado} high Curie temperature
($T_c$ $\approx$ 761\,K),~\cite{JPCM17_Markus_M4N} large magnetic
moment ($M_s$ $\approx$ 2.5\,\muB/Fe
atom)~\cite{Pssb09_Blanca_Fe4N} low coercivity and high chemical
stability.~\cite{Pssb09_Blanca_Fe4N} In addition, \tfn~possesses
an anti-perovskite structure which can be epitaxially grown on
most oxide substrates such as SrTiO$_3$, MgO and LaAlO$_3$.
\tfn~is also reported to exhibit the perpendicular magnetic
anisotropy with BiFeO$_3$~\cite{PRA16_PMA_Fe4N_BFO_Theo_Li} and
giant value of tunnel magneto resistance (TMR $\approx$ 24000\p)
with MgO based magnetic tunnel
junction.~\cite{PRA18_TMR_Fe4N_Yang_Theo} As a result,
\tfn~becomes an attractive candidate in spintronic devices due to
its TMR ratio, spin polarization ratio,
etc.~\cite{JAP09_Fe4N_75p_MR, APL09_Fe4N_SPR_60p_Narahara}

However, a large variation (from the theoretical value of
2.5\,\muB/Fe atom~\cite{Pssb09_Blanca_Fe4N}) can be seen in the
$M_s$ of \tfn~thin films studied so far (table~\ref{tab1}). In
some studies $M_s$ as high as 2.9\,\muB~\cite{APL08_Atiq_Fe4N} was
reported and in others as low as 1.3\,\muB/Fe
atom.~\cite{CRL15_Feng_Fe4N} In most other works the $M_s$ of
\tfn~was found between these two extremes shown in
table~\ref{tab1}. Such a large scattering in the values of $M_s$
can stem from various experimental factors: (i) measurement
accuracy (ii) phase purity and/or compositional variations across
the depth of the film and (iii) the deposition methodology
affecting the microstructure of the films. These factors need to
be sought.

It should be noted here that primarily bulk magnetization
measurements have been performed on \tfn~thin films in most of the
work hitherto which inherently includes the large foreseen errors
while estimating the film volume. On the other hand, studies on
the structural and magnetic depth profiling in \tfn~film are
completely lacking. Moreover, different deposition methodology may
also lead to different microstructure and hence different $M_s$
values in \tfn~thin film. Mostly, direct current magnetron
sputtering (dcMS)~\cite{APL08_Atiq_Fe4N, CRL15_Feng_Fe4N,
APL03_Fe4N_NbN_Loloee, JCG15_Wang_Fe4N, JMMM15_Dirba_Fe4N} and
molecular beam epitaxy (MBE)~\cite{PRB04_Fe4N_Costa,
APL11_Ito_Fe4N, JAP15_Ito_Fe4N_XMCD, APL09_Fe4N_SPR_60p_Narahara,
JAP09_Fe4N_75p_MR} methods have been extensively used to prepare
the \tfn~films. Whereas, relatively new but a very promising
technique - high power impulse magnetron sputtering (HiPIMS) has
not yet been employed. There are several advantages inherent to
the HiPIMS process over the conventional dcMS process such as
improvement of the film quality by denser microstructure and
enhanced adhesion etc.~\cite{JAP17_Anders_HMS,
JVST12_Gudmundsson_HMS} As compared to dcMS, in HiPIMS high-power
pulses are employed at low duty cycle ($<$
10\p)~\cite{JAP17_R_HiPIMS_Strijckmans} leading to enhanced
ionization of process gas and sputtered species. Therefore, the
fraction of ionized species exceeds neutrals. These unusual
properties of HiPIMS led to additional improvement in the film
quality.~\cite{JAP17_Anders_HMS, JVST12_Gudmundsson_HMS}

In the view of this, we scrutinize the factors affecting the
variation in $M_s$ in a systematic way in this work. We deposited
single phase and epitaxial \tfn~film on a LaAlO$_3$ substrate
(lattice parameter; LP = 3.79\,{\AA}) as it is almost
100\p~lattice matched with \tfn (LP =
3.79\,{\AA})~\cite{Pssb09_Blanca_Fe4N}. They were deposited using
three different techniques namely dcMS, N-plasma assisted MBE and
HiPIMS. We performed detailed depth profiling measurements on
these samples and found an interesting result that La from the LAO
and Fe from \tfn~interdiffuse at the film-substrate interface. The
extent of this interface gets affected due to differences in the
microstructure of samples grown using different methods. By
further performing magnetic depth profiling and element specific
magnetization measurements, we attempt to understand the role of
interface and microstructure in affecting the magnetization of
\tfn~thin films.

\begin{table}\center
\caption{\label{tab1} Magnetic moment ($M_s$) of \tfn~films
measured in different experimental works and theoretically
calculated using different methods [augmented plane waves plus
local orbitals (APW+lo), linearized muffin tin orbital method
(LMTO), atomic sphere approximation (ASA), full-potential
linearized augmented plane wave method (FLAPW), augmented
spherical wave method (ASW) with functions local density
approximation (LDA), generalized gradient approximation (GGA),
Perdew-Bruke-Ezerhof (PBE), Hubbard (U)]. Here Ref. denotes to
`Reference'.}
\begin{tabular}{ccccc}

\hline
&Experiemtnal Work&\\\\
\hline\hline

$M_s$&Technique&Ref.&\\
\muB/Fe&&&\\\hline

2.9&Bulk magnetisation&[\cite{APL08_Atiq_Fe4N}]&\\
1.3&Bulk magnetisation&[\cite{CRL15_Feng_Fe4N}]&\\
2.23&Bulk magnetisation&[\cite{JMMM15_Dirba_Fe4N}]&\\
2.47&x-ray circular magnetic dicorism&[\cite{APL11_Ito_Fe4N}]&\\
2.04&Bulk magnetisation&[\cite{APL03_Nikolaev_Fe4N_STO}]&\\
2.1&x-ray circular magnetic dicorism&[\cite{PRB10_Takagi_fe4n_XMCD}]&\\
2.12&Bulk magnetisation&[\cite{JMMM17_Golden_MBE_Fe4N}]&\\
1.65&Bulk magnetisation&[\cite{APL18_Fe4N_Cr_Ag_buffer}]&\\\\

\hline
&Theoretical Work&&\\\\
\hline\hline

$M_s$&Method/function&Ref.&\\
\muB/Fe&&&\\\hline

$>$2.4-2.8$<$&APW+lo/LDA+U, PBE+U&[\cite{Pssb09_Blanca_Fe4N}]&\\
2.29&LMTO+ASA/LDA&[\cite{JMMM91_Fe4N_Sakuma}]&\\
2.355&LMTO+ASA/LDA&[\cite{JMMM92_Kuhnen_Fe4N}]&\\
2.31&LMTO+ASA/LDA&[\cite{JPCM94_Fe4N_Fe16N2_Coey}]&\\
2.42&FLAPW/LDA&[\cite{PRB93_Coehoorn_Fe4N}]&\\
2.34&FLAPW/LDA&[\cite{JMMM99_Fe4N_Mohn_Matar}]&\\
2.16&FLAPW/LDA&[\cite{JMMM99_Fe4N_Sifkovits}]&\\
2.35&ASW/GGA&[\cite{JMMM10_Fe4N_Houari_Matar}]&\\

\hline\hline
\end{tabular}\\
\end{table}

\section{Experimental Procedure} \label{expe}
\tfn~films were grown on LaAlO$_3$ (100) substrate using
N-assisted MBE (DCA, M600 system at JCNS, Garching), direct
current magnetron (dcMS) and high power impulse mgnetron
sputtering (HiPIMS) (ATC Orion 8, AJA Int. Inc. at UGC-DAE CSR,
Indore) techniques. In MBE a rf N-plasma source was used to
evaporate Fe (99.995\p~pure) from an effusion cell. In MBE chamber
the base pressure was about 5$\times$10$^{-10}$\,Torr and with the
N$_2$ gas flow of 0.07 standard cubic centimeter per minute (sccm)
in the rf plasma source, the pressure during growth was about
1.3$\times$10$^{-7}$\,Torr. In dcMS and HiPIMS processes, Fe
targets (99.95\p~pure) - $\phi$~1\,inch and $\phi$~3\,inch were
used as a source, respectively. In HiPIMS process the peak power
was maintained at 26\,kW by keeping average power fixed at 300\,W,
peak voltage 700\,V, pulse frequency 75\,Hz and pulse duration
150\,$\mu$s. In dcMS process, the sputtering power was fixed at
5\,W. The partial gas flow of nitrogen ({\pn} =
p$_{\mathrm{N}_2}$/(p$_{\mathrm{Ar}}$+p$_{\mathrm{N}_2}$), where
p$_{\mathrm{Ar}}$ and p$_{\mathrm{N}_2}$ are gas flow of Ar and
N$_2$ gases, respectively) was kept at 10 and 23\p~for dcMS and
HiPIMS processes, respectively. A base pressure of
1$\times$10$^{-7}$\,Torr~was achieved before deposition and the
working pressure was maintained at 4\,mTorr~during deposition in
both dcMS and HiPIMS processes. The thickness of \tfn~samples was
about 50\,nm. An Au layer of thickness around 2\,nm was used as a
capping layer in MBE grown \tfn~film.

The crystal structure and the phase formation of the films were
characterized by x-ray diffraction (XRD) using a standard x-ray
diffractometer (Bruker D8 Advance) using CuK-$\alpha$ x-ray
source. Compositional depth profiling was performed using
secondary ion mass spectroscopy (SIMS) in a Hiden Analytical SIMS
workstation. A primary O$_2$$^+$ ions source was used for
sputtering with an energy of 3\,keV and beam current of 150\,nA.
The sputtered secondary ions were detected using a quadrupole mass
analyzer. X-ray reflectivity measurements were carried out using
using CuK-$\alpha$ x-ray source. Bulk magnetization measurements
were done using Quantum design SQUID-VSM magnetometer. Polarized
neutron reflectivity (PNR) measurements were performed at AMOR,
SINQ, PSI Switzerland in time of flight mode using Selene optics
on samples deposited using dcMS and HiPIMS.~\cite{Amor16_cite_1,
Amor17_cite_2} PNR measurements on MBE deposited samples were
carried out using the MAgnetic Reflectometer with high Incident
Angle (MARIA) of the JCNS, Garching, Germany.~\cite{2018_MARIA}
During PNR measurements, to saturate the sample magnetically, a
magnetic field of 0.5\,T was applied parallel to the sample
surface. X-ray magnetic circular dichroism (XMCD) measurement were
carried out at BL-01, Indus 2, RRCAT, India.~\cite{BL01_RRCAT} The
x-ray incidence angle was fixed at 90$^{\circ}$ with respect to
the sample surface. Magnetic anisotropy was studied using magneto
optical-Kerr effect (MOKE) and Kerr microscopy (Evico Magnetics)
equipment.

\section{Results and Discussion}
\subsection{\textbf{Structure and bulk magnetization}}

\begin{figure} \center
\vspace{-3mm}
\includegraphics [width=70mm,height=85mm]{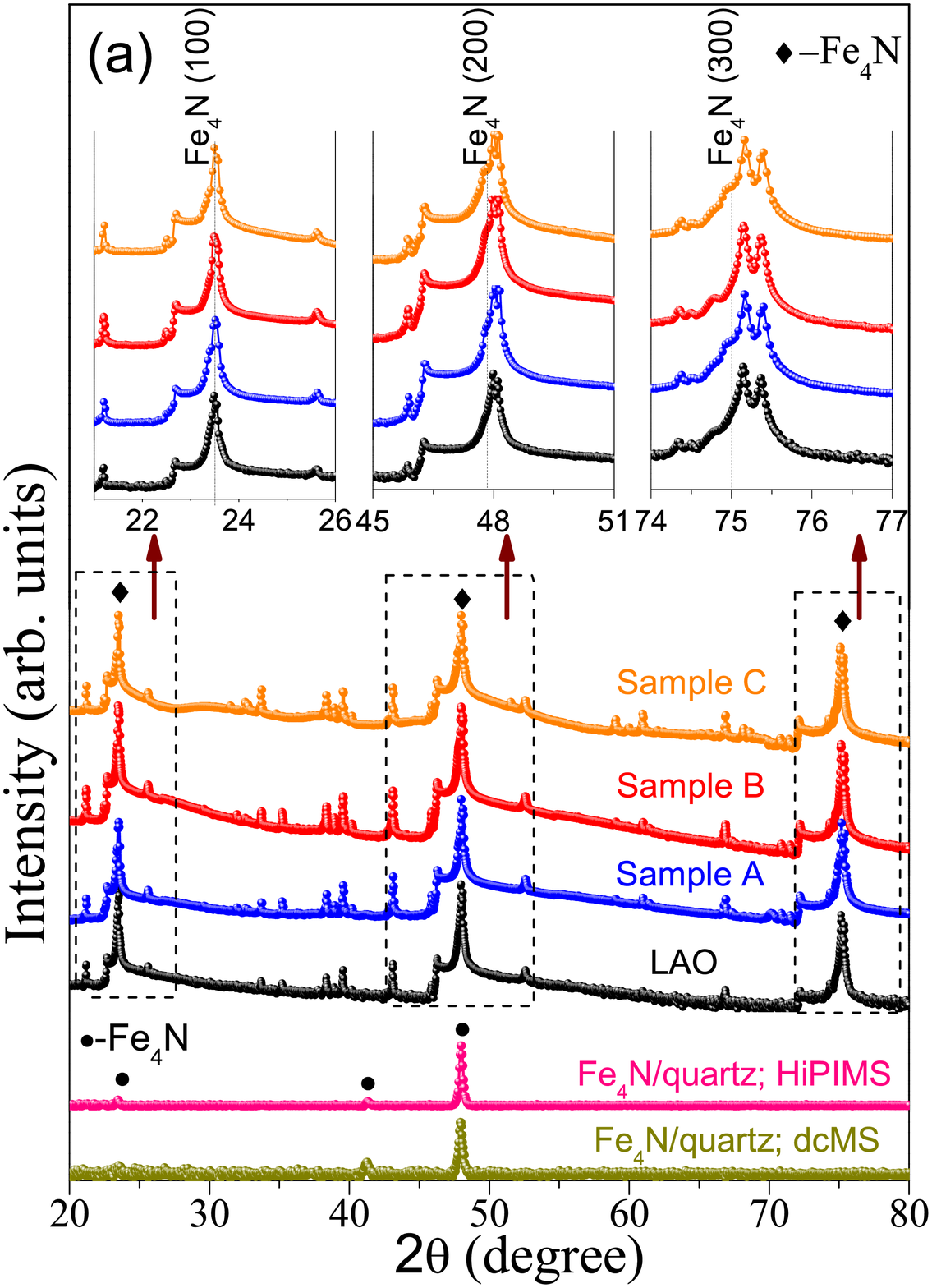}
\includegraphics [scale=0.62] {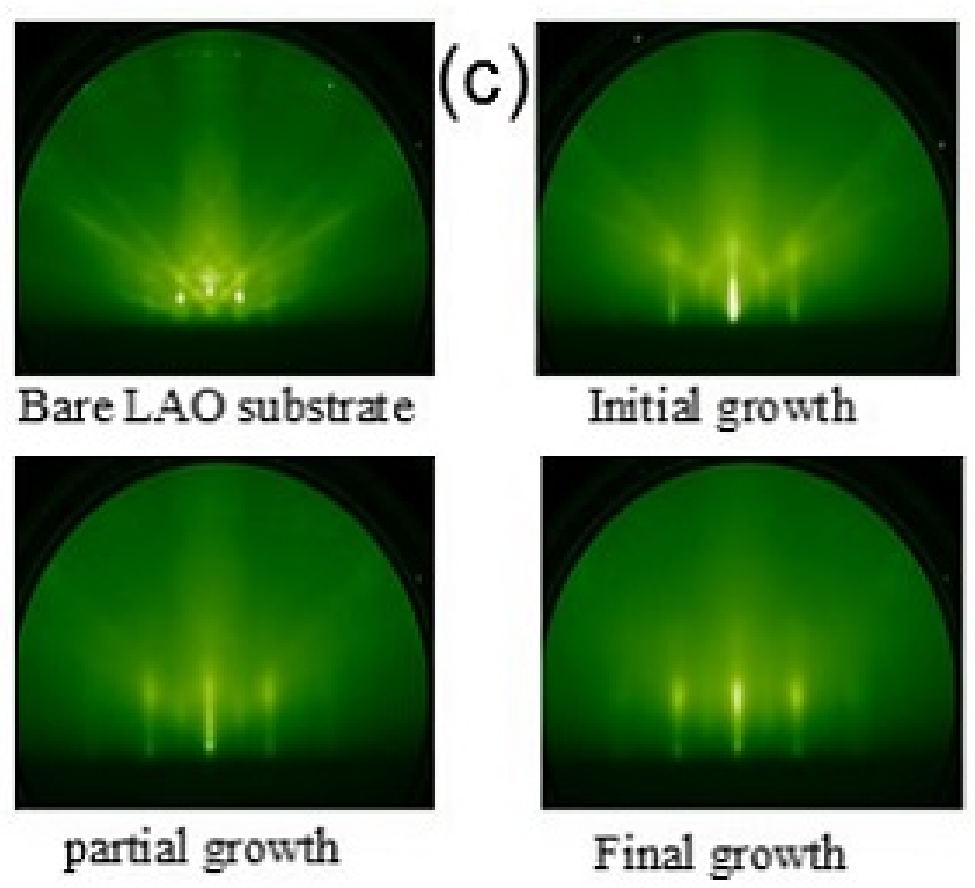}
\caption{\label{fig:xrd} XRD patterns of samples\,A, B and C along
with polycrystalline \tfn~film deposited on quartz substrate using
dcMS and HiPIMS and a bare LAO substrate (for reference) (a).
Inset showing the enlarge view of these XRD patterns shown by
arrow. RHEED images were captured during in-situ growth of
sample\,C at initial growth (4\,nm), partial growth (35\,nm) and
final growth (40\,nm) (b).} \vspace{-3mm}
\end{figure}

\tfn~thin film samples grown on LAO(001) substrate are labelled
as: dcMS (sample\,A), HiPIMS (sample\,B) and MBE (sample\,C) and
their XRD patterns are shown in fig.~\ref{fig:xrd} (a). For
reference, XRD patterns of two polycrystalline \tfn~films
deposited along with the above mentioned samples (grown on
amorphous quartz substrate using dcMS and HiPIMS) and a bare LAO
substrate are also included in fig.~\ref{fig:xrd} (a).
Polycrystalline samples demonstrated solely three peaks associated
to (100), (111) and (200) planes of \tfn~phase. This implies the
formation of single \tfn~phase. In addition, XRD patterns
(fig.~\ref{fig:xrd} (a)) of samples grown on the LAO substrate
shown only the reflections which are exactly matched with the LAO
substrate peaks. This could be understood from the fact that LAO
and \tfn~exhibit 0\p~lattice mismatching and therefore,
discrimination between the peak positions of LAO and \tfn~is not
possible. In order to distinguish the reflection of \tfn~phase, an
enlarged view corresponding to (100), (200) and (300) peaks has
been plotted shown in the inset of fig.~\ref{fig:xrd} (a). A
shoulder appeared towards lower 2$\theta$ can be seen in each case
confirming the presence of \tfn~phase. As expected, for higher
angle (300) plane, it is considerably noticeable rather than the
lower angle planes. Consequently, our XRD results confirms the
formation of single phased \tfn~film well-oriented along the c
axis (normal) of the substrate.

To confirm the epitaxial growth of our film, RHEED images were
taken in-situ during the growth of sample\,C and shown in
fig.~\ref{fig:xrd} (b). Here, the RHEED images of a bare LAO
substrate taken before and after several deposition times are
shown. During \tfn~growth streaks become visible proving epitaxial
growth. As an intensity modulation of the streaks evolves, we
conclude that the surface roughness increases due to terraces of
discontiguous widths.~\cite{Diffraction_Rheed_lagally1988} Hereby,
the RHEED images confirm the epitaxial nature of our \tfn~films.
Microstructure and surface morphology of these samples were
obtained using atomic force microscopy (AFM) measurements shown in
the Supplementary Material.~\cite{Suppl_Mater} From here, it can
be seen that HiPIMS grown samples exhibited a denser
microstructure and lower roughness.

\begin{figure} \center
\vspace{-1mm}
\includegraphics [scale=0.32] {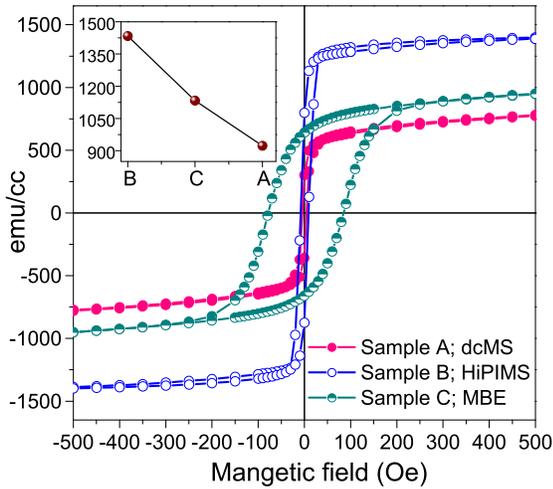}
\caption{\label{fig:mh} MH hysteresis curves of samples\,A, B and
C. Inset showing the respective $M_s$ values.} \vspace{-1mm}
\end{figure}

Bulk magnetization measurements were performed on all samples\,A,
B and C and corresponding MH hysteresis loops are shown in
fig.~\ref{fig:mh}. The coercivity ($H_c$) of dcMS and HiPIMS
samples is $\approx$ 20\,Oe but for sample\,C it is much higher
$\approx$ 100\,Oe. Such a higher value of $H_c$ for MBE grown
sample\,C may due to presence of different type of interfaces,
film-substrate and film-capping layer. This may lead to higher
pinning domain resulted in larger $H_c$. Moreover, even larger
difference can be seen in the values of saturation magnetization
($M_s$) for samples\,A, B and C shown in the inset of
fig.~\ref{fig:mh}. $M_s$ is the highest for sample\,B (HiPIMS)
while the lowest for sample\,A (dcMS). However, even the highest
obtained value of $M_s$ $\approx$ 1425\,emu/cc for HiPIMS grown
sample is still lower than its theoretical value $\approx$
1690\,emu/cc.~\cite{Pssb09_Blanca_Fe4N}

\subsection{\textbf{Structural and magnetic depth profiles of \tfn~films}}

From our XRD measurements, it can be seen that the structure of
samples grown using dcMS, HiPIMS and MBE is similar and confirm
the epitaxial growth of \tfn~on LAO substrate. But from our bulk
magnetization measurements and CEMS
measurements,~\cite{Suppl_Mater} overall values of $M_s$ are
smaller than the theoretically predicted
values.~\cite{Pssb09_Blanca_Fe4N} Differences in $M_s$ values can
also be seen in samples prepared using different techniques. In
order to understand such variances in $M_s$, we did elemental and
magnetic depth profiling using SIMS and PNR, respectively.

\begin{figure} \center
\vspace{-3mm}
\includegraphics [scale=0.3] {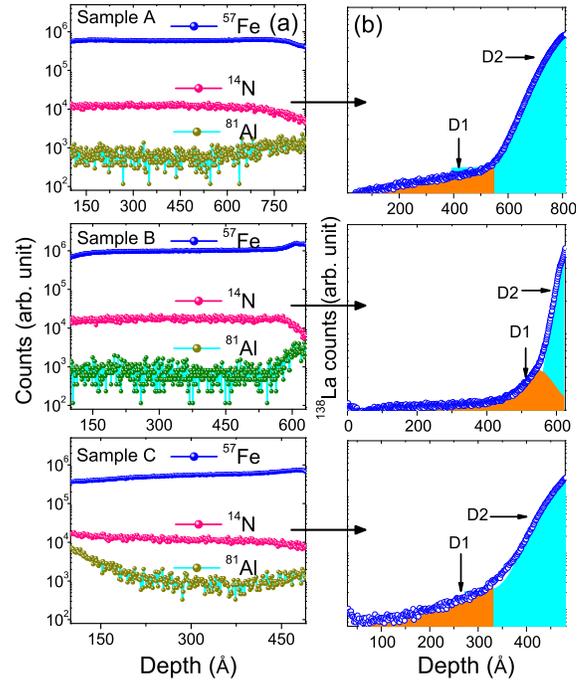}
\caption{\label{fig:sims} SIMS depth profiles of \tfn~thin films
deposited on LAO substrate using dcMS (sample\,A), HiPIMS
(sample\,B) and MBE (sample\,C) techniques. Fe, N and Al profiles
are shown in (a) respective La profiles (indicated by arrow) are
shown in (b).} \vspace{-1mm}
\end{figure}

SIMS depth profiles are shown in fig.~\ref{fig:sims} (a) for
samples\,A, B and C. Here, we can see that the Fe and N profiles
demonstrated nearly uniform behavior and an analogous distribution
to each other along the depth of the films in the samples\,A
(dcMS) and B (HiPIMS). Whereas, they seem to be skewed in
sample\,C (MBE) near to the surface and interface regions. This
reflects that the distribution of Fe and N is more uniform in
samples\,A and B while presence of some concentration gradient in
sample\,C. On the other hand, La depth profiles shown in
fig.~\ref{fig:sims} (b) reveal the mystery. They were deconvoluted
assuming two gaussian functions. Here, the component D2 can be
directly related to the film-substrate interface region and D1 to
an interdiffused region. The value of D2 comes out to be about 20,
7 and 14\,nm, for sample\,A, B and C, respectively reflects the
thinnest film-substrate interface in sample\,B (HiPIMS). These
results clearly evident that the La diffuses more into films grown
by dcMS and MBE and play a major role in forming a broad
interface. Similarly, the D1 contribution is quite profound in
sample\,A and C while and feeble in sample\,B. Therefore, it
appears that La propagated as an impurity into \tfn~quite
substantially in samples\,A and C but not so much in sample\,B.
Such interdiffusion has been previously probed in SrTiO$_3$/LAO
heterostructure and it was found that La forms a broader interface
(compared to Al in LAO) and has been described in terms of the
stability of LAO compound with oxygen
vacancies.~\cite{SurfSci11_Chambers_LAO/STO_diffusion,
SurfSci02_Kawanowa_LAO_diffusion} Oxygen depletion from LAO
induces the Al diffusion into sub surface regions but a change of
valency of La from 3$^+$ to 2$^+$ act as a driving force leading
to segregation of La to a much larger length
scales.~\cite{SurfSci02_Kawanowa_LAO_diffusion}

\begin{figure} \center
\vspace{-1mm}
\includegraphics [scale=0.29] {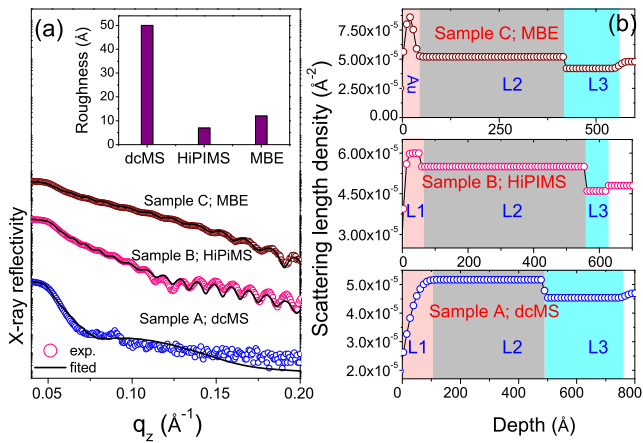}
\caption{\label{fig:xrr} XRR patterns (a) and SLD depth profiles
(b) of samples\,A, B and C. Inset of (a) showing the roughness of
samples\,A, B and C. Here, L1, L2 and L3 denoted the surface
region, bulk of the \tfn~film and film-substrate interface,
respectively.} \vspace{-1mm}
\end{figure}

To further confirm SIMS results, depth profiles were also obtained
from XRR measurements as shown in fig.~\ref{fig:xrr} (a). Fitting
of XRR data were performed (using Parraatt32~\cite{Parratt32})
considering a three layer model -(i) L1- surface region (ii) L2-
bulk of \tfn~film and (iii) L3- film-substrate interface. As shown
in fig.~\ref{fig:xrr} (b), the width of L3 is substantially small
in HiPIMS grown sample\,B as compared to samples\,A and C. This
behavior is in agreement with SIMS depth profiling results and the
width of the interface is also similar. As discussed before, such
variations can be understood due to larger La interdiffusion when
the microstructure is porous in dcMS and MBE grown samples but due
to denser microstructure, La diffusion gets suppressed leading to
sharper interface in HiPIMS grown sample. In addition, it can be
seen that surface roughness of dcMS grown sample is much higher in
agreement with AFM results shown in Supplementary
Material.~\cite{Suppl_Mater}

\begin{figure} \center
\includegraphics [scale=0.28] {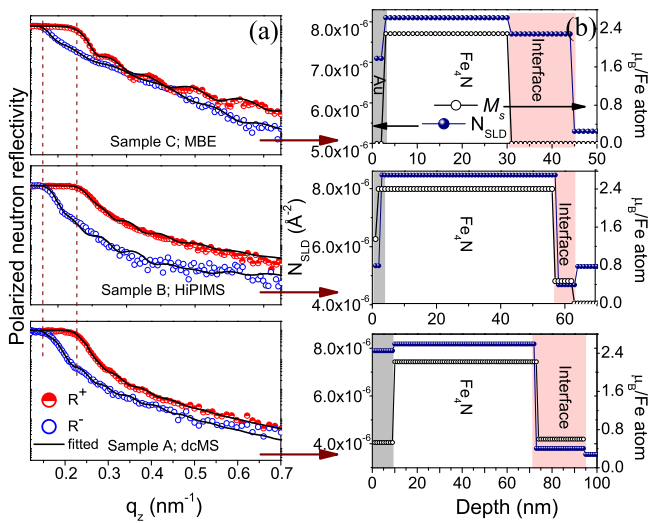}
\caption{\label{fig:pnr} Fitted PNR patterns (a) and corresponding
$\mathrm{N_{SLD}}$ and magnetic depth profiles (b) of samples\,A,
B and C shown by arrow.}
\end{figure}

The consequence of such film-substrate interface is also expected
to affect the magnetization behavior. Since the width of this
interface was lowest in HiPIMS grown sample, the value of
magnetization was largest. However, from bulk magnetization
measurements, contributions from interface layer can not be
separated. Therefore, we did PNR measurements in samples\,A, B and
C. It is well-known that the magnetic depth profile can be
uniquely obtained from PNR measurements but it was surprising to
note that they have not been performed in \tfn~thin films before.
Fig.~\ref{fig:pnr} (a) shows the PNR patterns for samples\,A, B
and C and they were fitted using GenX software.~\cite{GenX} It is
known that the splitting between spin up (R$^+$) and down (R$^-$)
neutron reflectivities near critical angle ($q_c$) is proportional
to the magnetization of the sample, given
by:~\cite{PNR92_Blundell}

\begin{equation}\label{L}
q_c{^\pm} = \sqrt{(16\pi N(b_n\pm b_m)}
\end{equation}

where, $N$ is the number density, $b_n$ and $b_m$ are the nuclear
and magnetic scattering lengths for neutrons. We can see that at
$q_c$ the separation between R$^+$ and R$^-$ is somewhat larger in
sample\,B, indicating higher $M_s$ in this sample. Taking inputs
from SIMS and XRR measurements, we again used a three layer model
described above and we can see that a film-substrate interface of
similar thickness was present in all three samples. From the
fitting of PNR data, we found that this interface layer is
magnetically dead as shown in fig.~\ref{fig:pnr} (b). The extent
of this layer was about 20, 7 and 15\,nm in dcMS, HiPIMS and MBE
grown samples, respectively. For the \tfn~layer (excluding surface
and interface) obtained values of $M_s$ = 1.8, 2.4 and 2.1 ($\pm$
0.05)\,\muB/Fe for sample\,A, B and C, respectively. This
difference in values of $M_s$ is in agreement with bulk
magnetization results. This value of $M_s$ matches well with the
theoretical value in sample\,B (see table~\ref{tab1}) but smaller
values in sample\,A and C can be understood due to the presence of
La impurity. Clearly, the microstructure and La diffusion affects
$M_s$ in \tfn~thin films. As can be seen from SIMS depth profiles,
La diffusion can prolong to a much larger length scale and thereby
affects the $M_s$. Since in HiPIMS grown film the La diffusion
could be suppressed due to a denser microstructure, the value of
$M_s$ reaches to the theoretical predicted value of
2.45\,\muB/Fe.~\cite{Pssb09_Blanca_Fe4N} Obtained results can be
applied to understand very large differences in the magnetization
of \tfn~films studied in the literature as shown in
table~\ref{tab1}. It can be anticipated that interdiffusion can
also take place from other substrates e.g. Si, SrTiO$_3$ and MgO
into \tfn~(or any other film) and in this situation the randomly
generated interface may lead to the randomness in the values of
$M_s$ that can be seen in \tfn~films grown in different
works.~\cite{APL11_Ito_Fe4N, PRB04_Fe4N_Costa, APL08_Atiq_Fe4N,
CRL15_Feng_Fe4N, APL03_Fe4N_NbN_Loloee, JCG15_Wang_Fe4N,
JMMM15_Dirba_Fe4N}

\subsection{\textbf{Elemental-specific magnetization}}

Theoretical calculations suggest a small but oppositely aligned
moment at the N site in \tfn. The origin of such magnetic moment
was explained in terms of the extension of spin-down electron wave
function near the interstitial region using spin-density plots
located within the muffin-tin
spheres.~\cite{JMMM99_Sifkovits_Nmoment,Coey_Mag_Magn_Mater10,APR16_review_Scheunert}
However, to the best of our knowledge, experimentally the magnetic
moment at the N site has only been studied by Ito $et~al.$ using
XMCD measurements,~\cite{JAP15_Ito_Fe4N_XMCD} but there also a
large discrepancy can be seen between the theoretically simulated
and experimentally observed N $K$-edge
spectra.~\cite{JAP15_Ito_Fe4N_XMCD} In the present case, as we
have shown that the \tfn~sample grown using HiPIMS was superior
and it will be interesting to inspect the electronic and magnetic
behavior at Fe and N sites.

\begin{figure} \center
\includegraphics [width=78mm,height=58mm] {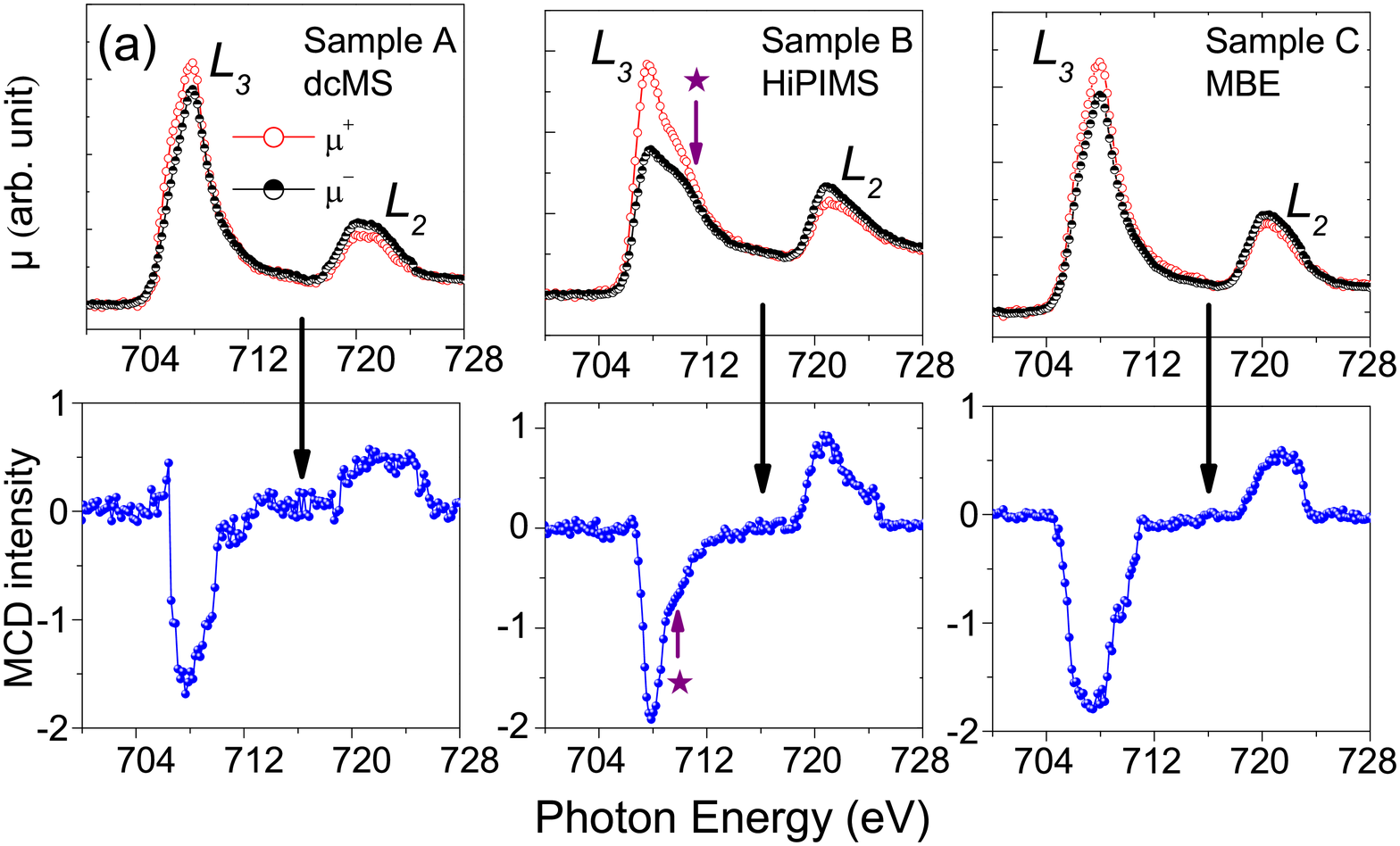}
\includegraphics [width=76mm,height=32mm] {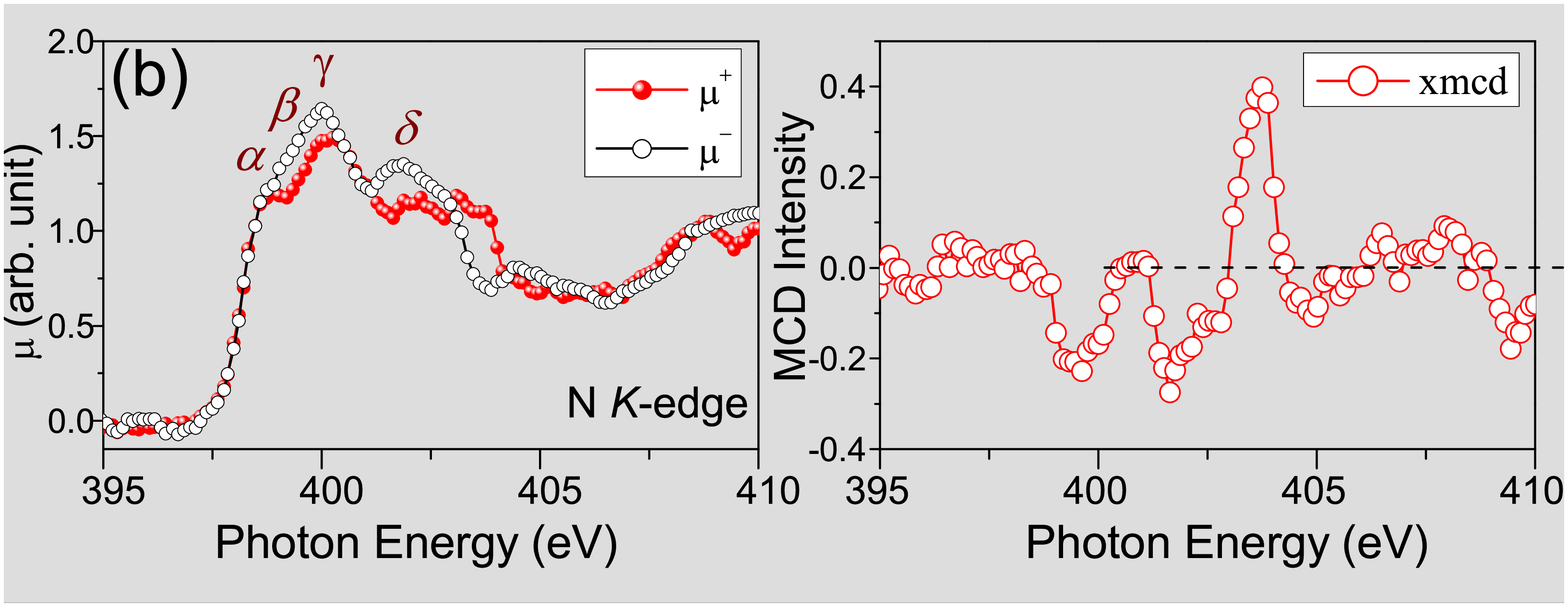}
\caption{\label{fig:mcd} XAS and XMCD spectra of samples\,A, B and
C observed at 300\,K at (a) Fe $L$-edge and (b) N $K$-edge. The
external magnetic field of $\pm$0.5\,T was applied along the x-ray
incidence direction to the sample surface.} \vspace{-1mm}
\end{figure}

\begin{table}\center
\caption{\label{tab2} Results of the sum-rule analysis of \tfn~
films (sample\,A, B and C). Average spin (m$_S$), orbital (m$_L$),
total (m$_{tot}$) magnetic moments, gyromagnetic ratio
(m$_L$/m$_S$) of Fe are given.}

\begin{tabular}{ccccc}\hline

Sample&m$_S$&m$_L$&m$_{tot}$&m$_L$/m$_S$\\
&\muB&\muB&\muB/Fe&\\
&$\pm$0.1&$\pm$0.05&&\\\hline\hline

A&1.56&0.07&1.64&0.046\\
B&2.23&0.13&2.36&0.058\\
C&1.86&0.07&1.92&0.042\\
\hline\hline
\end{tabular}\\
\end{table}

We performed, XAS and XMCD measurements at Fe $L$-edges at 300\,K
under UHV conditions by switching the applied external magnetic
field $\pm$0.5\,T ($\mu^+$ and $\mu^-$) along the direction of
propagation of x-ray beam leaving the x-ray helicity unchanged.
They are shown in fig.~\ref{fig:mcd} (a), here edges appearing at
photon energies of about 707 and 720\,eV can be seen and assigned
to Fe $L_3$ and $L_2$, respectively. A shoulder (denoted by
$\star$) can also be seen about 3\,eV above the $L_3$ edge and it
is more clearly pronounced in sample\,B grown using HiPIMS. It may
be noted here that such shoulder has been observed in some
metallic ferromagnetic systems and has been explained in terms of
unoccupied single-particle density of
states.~\cite{JAP15_Ito_Fe4N_XMCD} Here, this feature ($\star$)
can be attributed to the dipole transition from the Fe 2p
core-level to the hybridized state $\sigma^*$ between Fe (II)
sites~\cite{Suppl_Mater} and N 2p
orbitals.~\cite{PRB10_Takagi_fe4n_XMCD, APL11_Ito_Fe4N,
JAP15_Ito_Fe4N_XMCD} Since, this feature ($\star$) is noticeable
only in sample\,B, it indicates that the HiPIMS grown sample has
more localized states which could be due to better quality
\tfn~film. Distinct MCD spectra were observed at Fe $L$-edges in
all samples. Spin and orbital magnetic moments of samples\,A, B
and C were deduced by applying sum-rules analysis. The pre and
post-edge background correction was applied using Athena
software.~\cite{Athena} Transitions to the continuum states were
removed by subtracting the XAS average data using two step arc
tangent function.

It is known that, in sum-rules analysis, the magnetic moment is
proportional to the number of holes ($N_h$). In the present case,
we used $N_h$ = 3.88. This value was derived by Takagi
$et~al.$~\cite{PRB10_Takagi_fe4n_XMCD} for in-situ grown \tfn~thin
films on a Cu substrate. Obtained values of spin (m$_S$), orbital
(m$_L$) and total magnetic moments (m$_{tot}$) are shown in
table.~\ref{tab2} along with the gyromagnetic ratio (m$_L$/m$_S$)
for samples\,A, B and C. Our values of m$_L$/m$_S$ matched well
with the previously found values.~\cite{PRB10_Takagi_fe4n_XMCD,
APL11_Ito_Fe4N, JAP15_Ito_Fe4N_XMCD} Here, the lowest values of
total magnetic moments including orbital and spin magnetic moments
are found for the sample\,A (dcMS) while the highest for sample\,B
(HiPIMS), in agreement with bulk and PNR measurements.

We also did N $K$-edge XAS and XMCD measurements in sample\,B
(HiPIMS) as shown in fig.~\ref{fig:mcd} (b). Here, mainly four
features can be seen and they are assigned as $\alpha$, $\beta$,
$\gamma$ and $\delta$. The feature $\alpha$ is attributed to the
dipole transition from the N 1s to $\pi^*$ anti-bonding states and
features $\beta$ and $\gamma$ are explained by $\sigma^*$
anti-bonding states of N 2p and Fe 3d.~\cite{APL11_Ito_Fe4N,
JAP15_Ito_Fe4N_XMCD} Moreover, distinct XMCD spectrum is observed
at N $K$-edge confirms that N 2p orbital of \tfn~is
spin-polarized. It is also interesting to note that the $\mu^+$
and $\mu^-$ intensities get reversed compared to Fe $L_3$-edge
XMCD signal. This can be manifested considering an oppositely
aligned (negative) moment at the N site as compared to Fe as
predicted
theoretically.~\cite{JMMM99_Sifkovits_Nmoment,Coey_Mag_Magn_Mater10,APR16_review_Scheunert}
Also, our N $K$-edge XMCD spectra is well consistent with the
theoretically simulated spectra of Ito
$et~al$.~\cite{APL11_Ito_Fe4N}

\subsection{\textbf{Magnetic anisotropy}}

\begin{figure} \center
\vspace{-1mm}
\includegraphics [scale=0.22] {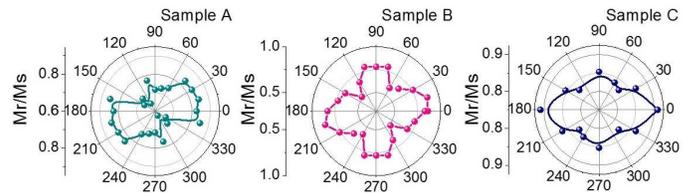}
\caption{\label{fig:moke} Polar plot of squareness (M$_r$/$Ms$)
with the applied field angle of samples\,A, B and C.}
\vspace{-1mm}
\end{figure}

We did longitudinal MOKE measurements to study magnetic anisotropy
(MA) and polar plots of reduced remanence (M$_r$/$M_s$; here M$_r$
and $M_s$ is the remanence and saturation magnetization) are shown
in fig.~\ref{fig:moke}. As can seen there, MA appears different in
these three samples. Only sample\,B (HiPIMS) demonstrated a
biaxial MA that is generally expected in cubic symmetry. Around
the easy magnetization axes (100) the reduced remanence is highest
close to 1(0.85) and around to the hard magnetization axes (110)
it is close to cos\,45$^{\circ}\simeq$~0.52. In addition, very
weak biaxial MA can be seen for sample\,C. On contrary, sample\,A
exhibited a small uniaxial MA. Unusual behavior of MA found in
samples\,A and C may be due to the significant diffusion of La
from the substrate to the film or due to substrate induce effect.
Here also, a discrepancy in the MA is clearly evident similar to
the magnetization of samples\,A, B and C. However, the detailed
investigation of such MA behavior needs to be further explored.

\begin{figure} \center
\vspace{-1mm}
\includegraphics [scale=0.38] {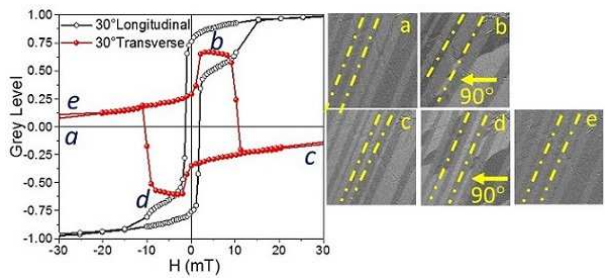}
\caption{\label{fig:kerr} Kerr domain images captured in between
easy and hard axis of sample\,B at 30$^{\circ}$ in transversal and
longitudinal directions.} \vspace{-1mm}
\end{figure}

As biaxial MA can only be seen in sample\,B, the magnetization
reversal by 90$^{\circ}$ domain is expected to appear in this
sample. Therefore, Kerr images were captured between easy and hard
axes for an applied field angle of 30$^{\circ}$ in transverse
direction. MH hysteresis loop of longitudinal and transverse
directions for applied field angle of 30$^{\circ}$ are shown in
the fig.~\ref{fig:kerr}. Images were captured for the transversal
MH loop at points $a(=e)$, $b$, $c$ and $d$. The cusp at points
$b$ and $d$ in both longitudinal and transversal directions
reflected the 90$^{\circ}$ domain wall driven transition.
180$^{\circ}$ magnetization reversal can be clearly seen from the
image $a$ to $e$ followed by two consecutive 90$^{\circ}$ domain
wall nucleation in image $b$ and $d$ (shown by arrow
90$^{\circ}$). However, stripe type domains called as lamellar
pattern can be seen in all images (shown by dashed lines in all
images). Such lamellar pattern domains originate as ferroelastic
domain arising due to occurrence of twin structures in
LAO.~\cite{Zaineb_2016_JMMM_LAO_Co}

\section{\textbf{Conclusion}}

In conclusion, we made an attempt to resolve the anomaly about
$M_s$ values of \tfn~thin films reported so far. In this view, we
have grown epitaxial \tfn~thin films on LAO substrate by utilizing
three different processes dcMS, MBE and HiPIMS and investigated
their structural and magnetic properties. $M_s$ of these samples
were measured using bulk magnetization, XMCD and PNR measurements.
Surprisingly, different $M_s$ values were found for all samples.
However, the highest value of $M_s$ was achieved for HiPIMS grown
sample. In order to probe the causes for the discrepancies in the
$M_s$ values, the detailed depth profiling and microstructure were
examined. A densely packed microstructure was found for HiPIMS
case. Whereas, a broad film-substrate interfaces were observed in
the \tfn~samples grown by MBE and dcMS compared to the HiPIMS
grown sample. The results obtained in this work directly signifies
that the denser microstructure may prohibits the La diffusion
inside the film resulting in a narrower film-substrate interface.
Our SIMS results elucidate that the $M_s$ in \tfn~highly
influenced by the La diffusion. Due to denser microstructure and
low film-substrate interaction in the HiPIMS grown sample, a small
magnetic moment at N atom was uniquely probed by element-specific
XMCD technique. In addition, the magnetic anisotropy behavior was
also found to be different in all samples. Four-fold magnetic
anisotropy expected for cubic symmetry was only observed for
HiPIMS grown sample. However, the origin of different kinds of
magnetic anisotropy requires a detailed investigation.

\section*{Acknowledgments}
NP is thankful to Council of Scientific Industrial Research (CSIR)
for senior research fellowship. Authors thank the Department of
Science and Technology, India (SR/NM/Z-07/2015) for the financial
support and Jawaharlal Nehru Centre for Advanced Scientific
Research (JNCASR) for managing the project. A part of this work
was performed at AMOR, Swiss Spallation Neutron Source, Paul
Scherrer Institute, Villigen, Switzerland. We acknowledge Zaineb
Hussain for fruitful discussion about MOKE measurements. We
acknowledge help received from L. Behera in sample preparation and
various measurements. We are thankful to V. Ganesan and M.
Gangrade for AFM measurements, R.J. Choudhary for S-VSM
measurements, R. Sah and A. Wadikar for XMCD measurements.

\section*{References}

%

\end{document}